\documentclass[aps,prb,superscriptaddress,twocolumn,nopacs,letter]{revtex4-2}
\usepackage{graphicx,bm,times}
\usepackage{amsmath}
\usepackage{amsfonts}
\usepackage{amssymb}
\usepackage{color}
\usepackage{hyperref}
\hypersetup{
	colorlinks = true,
	allcolors = {blue}
}

\usepackage[utf8]{inputenc}
\usepackage[T1]{fontenc}
\usepackage{graphicx}
\usepackage{amsmath,amssymb}
\usepackage{lineno}
\usepackage{float}
\usepackage{siunitx}

\begin{document}

\title{Fermiology and electron-phonon coupling in the 2H and 3R polytypes of NbS$\boldsymbol{_2}$}

\author{Zakariae El Youbi}
\affiliation {Diamond Light Source, Harwell Campus, Didcot, OX11 0DE, United Kingdom}
\affiliation {Laboratoire de Physique des Matériaux et des Surfaces, CY Cergy Paris Université, 5 mail Gay-Lussac, 95031 Cergy-Pontoise, France}
\author{Sung Won Jung}
\affiliation {Diamond Light Source, Harwell Campus, Didcot, OX11 0DE, United Kingdom}
\author{Christine Richter}
\affiliation {Laboratoire de Physique des Matériaux et des Surfaces, CY Cergy Paris Université, 5 mail Gay-Lussac, 95031 Cergy-Pontoise, France}
\affiliation{Université Paris-Saclay, CEA, CNRS, LIDYL, 91191, Gif-sur-Yvette, France}
\author{Karol Hricovini}
\affiliation {Laboratoire de Physique des Matériaux et des Surfaces, CY Cergy Paris Université, 5 mail Gay-Lussac, 95031 Cergy-Pontoise, France}
\affiliation{Université Paris-Saclay, CEA, CNRS, LIDYL, 91191, Gif-sur-Yvette, France}
\author{Cephise Cacho}
\affiliation {Diamond Light Source, Harwell Campus, Didcot, OX11 0DE, United Kingdom}
\author{Matthew D. Watson}
\email{matthew.watson@diamond.ac.uk}
\affiliation {Diamond Light Source, Harwell Campus, Didcot, OX11 0DE, United Kingdom}

\date{\today}

\begin{abstract}

We investigate the electronic structure of the 2H and 3R polytypes of NbS$_2$. The Fermi surfaces measured by angle-resolved photoemission spectroscopy show a remarkable difference in size, reflecting a significantly increased band filling in 3R-Nb$_{1+x}$S$_2$ compared to 2H-NbS$_2$, which we attribute to the presence of additional interstitial Nb which act as electron donors. Thus we find that the stoichiometry, rather than the stacking arrangement, is the most important factor in the difference in electronic and physical properties of the two phases. Our high resolution data on the 2H phase shows kinks in the spectral function that are fingerprints of the electron-phonon coupling. However, the strength of the coupling is found to be much larger for the the sections of bands with Nb 4$d_{x^2-y^2,xy}$ character than for the Nb 4$d_{3z^2-r^2}$. Our results provide an experimental framework for interpreting the two-gap superconductivity and "latent" charge density wave in 2H-NbS$_2$.  

\end{abstract}

\maketitle

\section{Introduction}

The transition metal dichalcogenides (TMDs) are well-known for hosting a variety of instabilities arising from the interplay of electron-electron and electron-phonon coupling. Particularly rich phenomena are found in the metallic (V,Nb,Ta)(S,Se,Te)$_2$ family, including Mott-insulating phases, superconductivity, and numerous charge density waves (CDW) \cite{Wilson1974PRL,Sipos.Nat.Mat.2008,RossnagelReview2011,Ritschel2015}. As a well-known example, 2H-NbSe$_2$ exhibits a $\sim{}3\times{}3$ CDW and also superconducts at 7.2 K \cite{Rossnagel.PRB.2012,Valla.PRL.2004,Borisenko.PRL.2009,Flicker.Nat.Comms.2015}. Several of these layered van der Waals materials favor the trigonal prismatic coordination of the transition metal, but there is an additional degree of freedom in the inter-layer stacking pattern (e.g. 2H, 4H, 3R polytypes \cite{Wilson.Adv.Phys.1969}), leading to further variety of the novel electronic ground states.

Unlike most other members of the (V,Nb,Ta)(S,Se,Te)$_2$ family, 2H-NbS$_2$ does not undergo any structural instability \cite{Bahramy.Nat.Comms.2020}. Nevertheless, a phonon mode exhibits significant softening with temperature \cite{Leroux.PRB.2012}, and 2H-NbS$_2$ can be viewed as being close to a lattice instability. This presents an interesting theoretical challenge, as na\"ive Density Functional Theory (DFT) calculations would predict a lattice instability \cite{Bianco.Nano.Letters.2019}, and the absence of any CDW phase is attributed to the anharmonic phononic effects \cite{Leroux.PRB.2012,Feliciano.PRL.2017}.  2H-NbS$_2$ is also a prototypical two-gap superconductor below T$_c$ = 6.2~K, as evidenced by tunneling spectroscopy \cite{Guillamon.PRL.2008}, Andreev reflections \cite{Majumdar2020}, and heat capacity measurements \cite{Kacmarcik.PRB.2010}. Despite this, the experimental electronic structure of bulk 2H-NbS$_2$ has hardly been explored \cite{Mannella.PRB.2016}. Meanwhile, an alternative stacking structure, the non-centrosymmetric 3R phase, has been reported for some time \cite{Fisher.Inorg.Chemi.1980,Meerschaut.Mat.Res.Bull.2001}, but little is known about its electronic structure, or how and why its properties differ from the 2H phase.

Here, we examine the low-energy electronic structure of 2H-NbS$_2$ and 3R-Nb$_{1+x}$S$_2$ combining Angle-Resolved Photoemission Spectroscopy (ARPES) and DFT calculations. The measured Fermi surfaces reveal a striking difference in size, implying a significantly greater band filling in the 3R phase. We attribute this difference to additional Nb interstitials in the 3R phase, which act as electron donors. This difference is likely to move the 3R phase away from any latent instabilities and may explain the absence of superconductivity in this phase. Finally, the high-resolution data on the 2H phase reveals that the electron-phonon coupling is highly dependent on the orbital character of the bands, which naturally links to the two-gap superconductivity. 

\section{Methods}
Single crystals were obtained commercially (HQ Graphene, Groningen) and cleaved \textit{in-situ}. ARPES measurements were performed at the I05 beamline of Diamond Light Source \cite{HoeshRev.Sc.Instrum2017}, using photon energies in the range 30-240 eV, at sample temperatures below 10 K. DFT calculations were performed within the Wien2k package \cite{wien2k2020}, using the modified Becke-Johnson (TB-mBJ) functional \cite{KollerPRB2012} and accounting for spin-orbit coupling, as further detailed in the Supplemental Material (SM, \cite{SM}). 

\begin{figure*}
	\centering
	\includegraphics[width=1\textwidth]{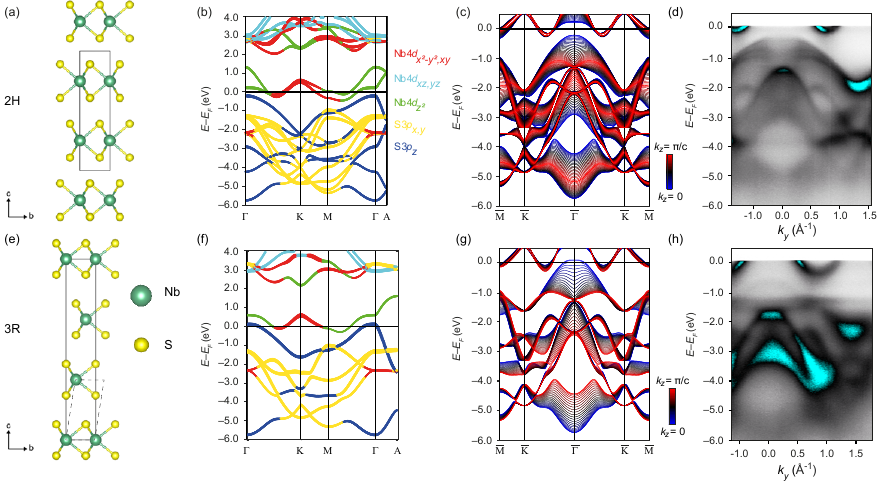}
	\caption{{Crystalline and electronic structure of 2H-NbS$_2$ and 3R-NbS$_2$.} (a,e) Crystal structures, showing the different stacking modes. (b,f) DFT calculations with orbital character projection of the valence and conduction bands for 2H-NbS$_2$ (b) and 3R-NbS$_2$ (f). (c,g) $k_z$ projection of the DFT band structure along the experimentally-relevant M-K-$\Gamma$-K-M (L-H-A-H-L) direction for 2H-NbS$_2$ (c) and 3R-NbS$_2$ (g). (d,h) Overview ARPES spectra showing valence band dispersions. Data measured along $\Gamma$-K direction at $h\nu =$ 79 eV and $h\nu =$ 68 eV for 2H-NbS$_2$ (d) and 3R-NbS$_2$ (h), respectively, using LH ($p$) polarized light.}
	\label{fig.1}
\end{figure*}

\section{Results}
\subsection{Comparison of the 2H and 3R phases}

In the 2H phase (space group 194, $P6_3/mmc$), each NbS$_2$ layer is rotated by $\ang{180}$ with respect to the layer below it (Fig. \ref{fig.1}(a)). Due to the trigonal prismatic coordination of the Nb, a single layer would not possess inversion symmetry, however in the 2H phase a centre of inversion symmetry exists between the layers \cite{Yuan.Nat.Phys.2013,Zhang.Nat.Phys.2014,Riley.Nat.Phys.2014}. Contrastingly, in the 3R phase  (space group 160, $R3m$) shown in Fig. \ref{fig.1}(e) there is no rotation but rather each layer is translated by a third of a lattice constant in the $b$ direction, with respect to the layer below. This stacking structure does not contain any points of inversion.

In our DFT calculations in Fig.~\ref{fig.1}(b,f), we find twice as many bands in the 2H phase compared to the 3R phase, since the 2H unit cell contains two formula units, while  the 3R phase contains one (in the primitive unit cell). The height of the Brillouin zone (i.e. $\Gamma$-A) in the 2H phase is half that of the 3R phase. Around the K point, in the 2H phase the bands near $E_F$ with Nb $4d_{x^2-y^2,xy}$ character (red) are split due to combination of interlayer hopping and the spin-orbit coupling, as discussed in numerous studies of 2H-MoS$_2$ \cite{Suzuki.Nat.Nano.2014,AlidoustNcomms2014,Razzoli.PRL.2017}, 2H-WSe$_2$ \cite{Riley.Nat.Phys.2014,Yuan.NanoLett.2016}, and 2H-NbSe$_2$ \cite{Bawden.Nat.Comms.2016}. However the Nb $4d_{3z^2-r^2}$ orbital (green) is more significantly affected by interlayer hopping terms, leading to a large band splitting of $\sim$1~eV at the $\Gamma$ point above $E_F$ in Fig.~\ref{fig.1}(b). For the in-plane S $3p_{x,y}$ valence bands (yellow), we find a small splitting of the band dispersions in the 2H phase compared to the 3R, since for these orbitals the interlayer hopping terms are relatively weak compared to the in-plane hoppings. In contrast, the S $p_z$ orbitals (blue) disperse strongly in the out of plane directions, and are most strongly sensitive to the stacking sequence.

In both phases, therefore, there are both quasi-2D and rather 3D valence bands, as highlighted in the $k_z$-projection of the band structure (Fig. \ref{fig.1}(c) and \ref{fig.1}(g)). This is a helpful representation of the DFT band structure for comparison with ARPES measurements, due to the non-conservation of $k_z$ in the photoemission process, which leads to an effective integration over a range of $k_z$ values \cite{Strocov2003,StrocovPRB2006,Riley.Nat.Phys.2014,ElYoubi.PRB.2020}. In the overview ARPES spectra of the 2H phase in Fig. \ref{fig.1}(d), the valence bands closely resemble the $k_z$-projected calculations, including sharp features from quasi-2D states, and broad features from 3D bands. In the case of the 3R phase, however, the agreement is notably worse. First, the data quality in Fig.~\ref{fig.1}(h) is lower; none of the features are as sharp as in the 2H phase, with a significantly higher background signal. Second, there is evidence for a flat state at $E_B$ = -1.2 eV, not present in the calculations, which is indicative of some form of localised impurity-like state. Third, the Nb $4d$-derived bands at the Fermi level have a substantially larger filling than in the calculation. 

\begin{figure*}
	\centering
	\includegraphics[width=1\textwidth]{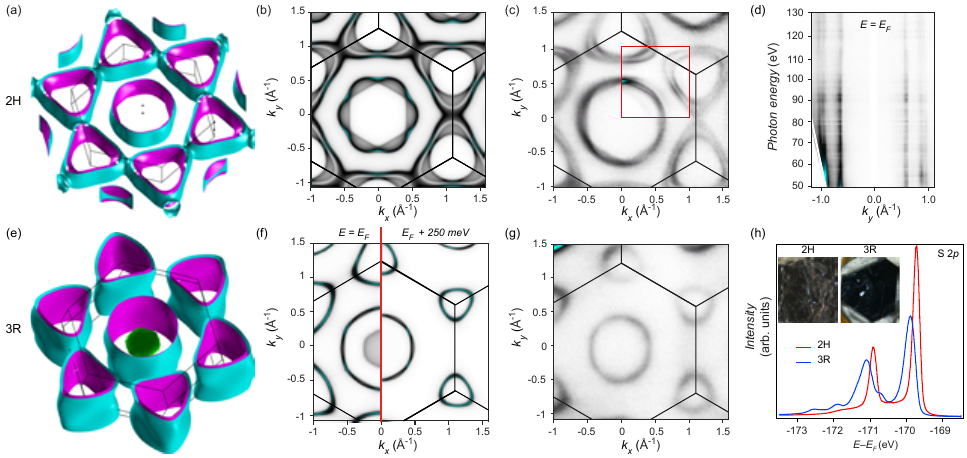}
	\caption{{Fermiology of 2H-NbS$_2$ and 3R-NbS$_2$.} (a,e) calculated 3D Fermi surface of (a) 2H-NbS$_2$ and (e) 3R-NbS$_2$. (b,f) Simulated Fermi surfaces obtained by averaging over the whole Brillouin zone width in the $k_z$ direction for 2H-NbS$_2$ (b) and 3R-NbS$_2$ (f). In the latter, a section is simulated at 250 meV above the natural Fermi level. (c,g) Fermi surface measured at a photon energy $h\nu =$ 79 eV (inset, 42 eV) for 2H-NbS$_2$ (c), and $h\nu =$ 120 eV  for 3R-NbS$_2$ (g). (d) Photon energy-dependent ARPES of 2H-NbS$_2$ from 50 to 130 eV, plotting MDCs at $E_F$ along K-$\Gamma$-K (H-A-H) direction as a function of photon energy, showing the quasi-2D nature of electronic states with a consistently-resolved splitting of the K barrel bands. (h) S 2$p$ core levels of 2H-NbS$_2$ (red) and 3R-NbS$_2$ (blue) using a photon energy $h\nu =$ 240 eV, showing clearly additional satellites in the 3R phase. Inset shows microscope images of the single crystals used in this work.}
	\label{fig.2}
\end{figure*}

To further understand the difference between the two phases, in Fig.~\ref{fig.2} we consider the Fermi surfaces. The DFT calculations of the Fermi surfaces in the two cases are broadly similar, as in both cases quasi-2D "barrels" appear, centered around the $\Gamma$ and K points (Fig. \ref{fig.2}(a) and \ref{fig.2}(e)). The inter-layer hopping in the 2H phase plays an important role in creating a splitting of inner and outer barrels around K. However there is no such splitting term in the 3R phase, and the splitting is a spin-splitting, allowed due to the absence of inversion symmetry \cite{SM}. In Fig. \ref{fig.2}(b) and \ref{fig.2}(f) we present simulations of the in-plane Fermi surfaces after averaging over the entire $k_z$ axis  \cite{2H3Rgeometry}. The measured Fermi surface in the 2H phase, Fig. \ref{fig.2}(c), is broadly comparable with the calculation. Notably, the band splitting around the K barrels is reproduced, and we can also resolve the two separate bands forming the $\Gamma$ barrel, with the inner displaying a strong hexagonal warping. The most noticeable difference is that experimentally, the triangular barrels around K form closed pockets (similar to 2H-NbSe$_2$ \cite{Borisenko.PRL.2009,Bawden.Nat.Comms.2016}),  while  in our calculation the outer K barrels connect near the M points. We attribute this to a limit to the accuracy of the functional, rather than any off-stoichiometry of the 2H sample. The more advanced GW calculations of Ref.~\cite{Feliciano.PRB.2018} similarly found smaller, closed, K barrels, along with slightly expanded $\Gamma$ barrels (see SM for detailed comparison \cite{SM}). Experimentally, we find that the observed Fermi surface appears highly two-dimensional, with very little variation observed in the photon energy dependence in Fig.~\ref{fig.2}(d).

However, in the measured Fermi map of 3R-NbS$_2$ (Fig. \ref{fig.2}(g)), the $\Gamma$ and K barrels of the Fermi surface are found to be both significantly smaller compared to the calculations Fig.~\ref{fig.2}(f, $k_x<0$). This is consistent with the increased filling of the band in Fig.~\ref{fig.1}(h), and is indicative of a large shift in the chemical potential. Instead, the data closely resembles the simulation in Fig.~\ref{fig.2}(f, $k_x>0$) where the Fermi level is set 250 meV above the natural Fermi level of the calculation. This implies a significant amount of extra charge in the system; in the calculation, a rigid shift of 250 meV as shown corresponds to 0.483 extra electrons per unit cell.

\begin{figure*}
	\centering
	\includegraphics[width=1\textwidth]{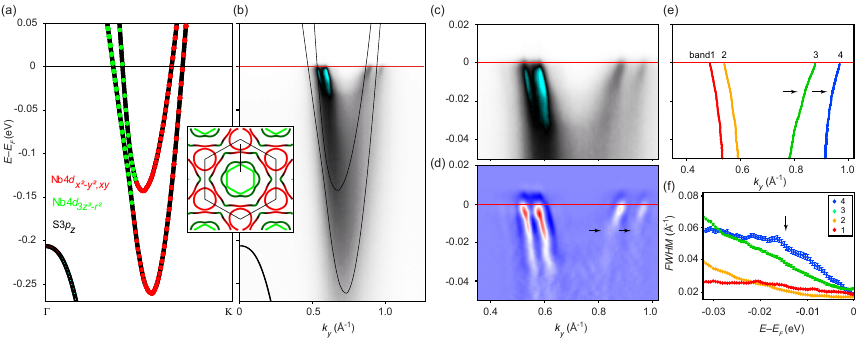}
	\caption{{Electron-phonon coupling in 2H-NbS$_2$.} (a) Calculated band dispersion along $\Gamma$K direction, with orbital character projection. Inset: calculated 2D Fermi surface; the thick black line indicates the experimental cut analysed in (c-f). (b) ARPES data of the valence band dispersion along $\Gamma$K, measured at a photon energy $h\nu =$ 30 eV, overlaid with DFT calculations for comparison. (c) Closer look at the data near $E_F$ and (d) the `curvature' plot of the data, highlighting kinks in the spectral function. (e) Momentum distribution curve (MDC) fitting of the data using a multi-Lorentzian peak function, and (f) peak widths of the four bands.}
	\label{fig.3}
\end{figure*}

The experimental data therefore point toward an important role for the stoichiometry in determining the electronic structure differences between the two polytypes. While the 2H phase is reported to exist only as stoichiometric NbS$_2$, the 3R phase is known to host additional Nb interstitials, resulting in a stoichiometry of the form 3R-Nb$_{(1+x)}$S$_2$. According to Ref. \cite{Fisher.Inorg.Chemi.1980}, the range of stability of the 3R phase is 0.07 < $x$ < 0.18; consistent with this, energy-dispersive X-ray spectroscopy (EDX) measurements on our 3R samples indicate $x\approx{}$0.13 \cite{EDX}. According to Ref. \cite{Meerschaut.Mat.Res.Bull.2001}, interstitial atoms are more favorably incorporated in the 3R phase due to the relatively longer distance between these extra Nb atoms and the in-plane Nb sites. Thus, the fact that there exists two polytypes at all is intimately related to the stoichiometry. These Nb interstitials act as electron donors, giving an increased overall filling of the $d$ shell of the Nb in the main layers. Taking the ratio of the apparent extra electrons in the Fermi surface to $x$, we can estimate that each interstitial Nb donates $\sim$ 3.7 electrons on average. The interstitial sites also act as local impurity potentials, explaining the broadening and extra background in the ARPES data in e.g. Fig.~\ref{fig.1}(h). 

The presence of interstitial Nb can also be inferred from the S 2$p$ core level spectra in Fig~\ref{fig.2}(h). In the case of the 2H phase, a sharp doublet is observed, consistent with spin-orbit split $2p_{1/2}$ and $2p_{3/2}$ states from a single chemical site. However, in the 3R phase, there are additional minority peaks, arising from S atoms in a chemical environment with more than the normal three nearest-neighbour Nb atoms, due to the interstitial Nb occupancy. The main doublet is also broadened, reflecting electronic inhomogeneity caused by the partial filling of the interstitial sites. Moreover, the main doublet is shifted by $\sim$ 180 meV, a chemical shift related to the overall chemical potential and average orbital fillings. It is worth remarking that the two polytypes also have different physical appearances; although both black and metallic-looking, the off-stoichiometric 3R phase forms as beautiful plate-like single crystals with clear crystal facets, while the 2H phase forms thin, flaky samples usually without clear facets. Thus, NbS$_2$ exemplifies the mantra of not judging by appearances.

\subsection{Electron-phonon coupling in the 2H phase}

Although neither phase undergoes any structural phase transition, 2H-NbS$_2$ is considered to be on the brink of a CDW-like transition \cite{Guillamon.PRL.2008,Kacmarcik.PRB.2010,Diener.PRB.2011,Leroux.PRB.2012,Tissen.PRB.2013,Feliciano.PRL.2017,Bianco.Nano.Letters.2019,Wen.PRB.2020}, while a CDW does stabilise in the closely-related 2H-NbSe$_2$ at 33.5 K. It is generally acknowledged that the charge density waves in this family of materials cannot be explained by electronic ``nesting" alone, and it is important to consider the momentum-dependence of the electron-phonon interaction, which itself is related to the orbital character of the bands \cite{Rossnagel.PRB.2001,Johannes.PRB.2006,Johannes.PRB.2008,Borisenko.PRL.2009,Rossnagel.PRB.2012,Flicker.Nat.Comms.2015,Flicker.PRB.2016}. For the calculated bands along $\Gamma$K in Fig.~\ref{fig.3}(a), there is a crossover in orbital character between the doublets corresponding to the $\Gamma$ barrels (mainly Nb 4$d_{3z^2-r^2}$) and the K barrels (mainly Nb 4$d_{x^2-y^2,xy}$, see also Fig.S2 \cite{SM}). The experimental spectral function in Fig.~\ref{fig.3}(b) shows significantly different impact of electron-phonon coupling at the two pairs of Fermi crossings, with the second pair of crossings (corresponding to the K barrels) exhibiting a clear kink structure, characteristic of a strong electron-phonon interaction, whereas for the first pair ($\Gamma$ barrels) the effect is much less prominent. 

For a more quantitative analysis, we performed a fitting analysis to extract the band positions in Fig.~\ref{fig.3}(e). From this, we identify a ``kink" energy of -15 meV for the K barrels; the deviation of the bands around this energy can also be visualised in the ``curvature" plot in Fig.~\ref{fig.3}(d). Meanwhile the inner bands have a change of slope around -20 meV, but this is a more subtle effect. A quantitative measure of electron-phonon coupling is the renormalization of the Fermi velocity, $\lambda=v_F(\mathrm{bare})/v_F(\mathrm{exp.})-1$ \cite{Hofmann.New.J.Phys.2009}. If we assume the $v_F(\mathrm{bare})$ values from DFT (see SM for discussion \cite{SM}), we find a clear dichotomy between the innermost band 1, with $\lambda{}\approx{}0.51$, and the outer crossings of the K barrel with $\lambda{}\approx{}2.32$ for band 3 and $\lambda{}\approx{}2.59$ for band 4. Additionally, in Fig.~\ref{fig.3}(f) we show a dichotomy in the energy-dependent linewidths, as the broadening of the K barrels increase much faster with binding energy than the $\Gamma$ barrels. The outermost band 4 shows the fastest rise, consistent with having the strongest coupling, and also shows a saturation of the linewidth coinciding with the kink energy \cite{Hofmann.New.J.Phys.2009}.

Taken together, this evidence strongly suggests that in 2H-NbS$_2$, the electron-phonon coupling depends crucially on the orbital character of the bands, and for the $\Gamma$K dispersion analysed here, the $\lambda{}$ value is up to $\sim$ 4-5 times larger for the section with Nb 4$d_{x^2-y^2,xy}$ character than for the Nb $4d_{3z^2-r^2}$. Our experiments are highly consistent with the calculations of Ref.~\cite{Feliciano.PRL.2017}, who took the electron-phonon interaction into consideration and also found a significantly larger interaction strength on the sections of the K barrel closest to the K points, correlating closely with the Nb 4$d_{x^2-y^2,xy}$ orbital character. This has important implications for the superconductivity, and our data gives strong experimental support for the scenario of Ref.~\cite{Feliciano.PRL.2017}, where the inner $\Gamma$ sheets with weaker el-ph coupling are "cold" areas corresponding to the smaller gap, while the straight sections of the K barrels are "hot" regions developing a larger gap, explaining the overall gap structure with two characteristic energy scales \cite{Guillamon.PRL.2008,Majumdar2020}. 

\section{Discussion}

Recapping the results on 3R-Nb$_{(1+x)}$S$_2$, we showed that the prevalence of donor-type interstitials leads to a significant chemical potential shift by approximately 250 meV, compared to either the calculations for stoichiometric 3R-NbS$_2$, or compared to the 2H phase. This implies an extra population of the Nb $4d$ bands that dominate the Fermi surface, an increased filling of $\sim$0.48 extra electrons per Nb. The resulting Fermi surface pockets are still hole-like but substantially smaller in area, and calculations indicate a reduced electronic density of states (DOS) at $E_F$ by a factor of $\sim$2 at this doping level (SM). A reduced DOS in the 3R phase is, furthermore, consistent with a magnetic susceptibility that is lower by a factor of 2-3 in the 3R phase compared with the 2H phase, in the Pauli paramagnetic regime \cite{Fisher.Inorg.Chemi.1980}. While a full treatment would also also take into account differences in phononic structure between the two stacking sequences, the reduced electronic DOS alone is already likely to move the doped 3R system further away from the latent structural instabilities found in the 2H phase \cite{Leroux.PRB.2012}. Similarly, in terms of superconductivity, the lower electronic DOS in the 3R phase will be a factor in lowering $T_c$, to the point that no superconductivity has yet been reported in the 3R phase. The calculations using Eliashberg theory in Ref.~\cite{Nishio.JPSJ.1994} found a reduction of $T_c$ by at least a factor of 3 for a positive chemical potential shift of 150 meV in 2H-NbS$_2$, while our 3R phase sample is effectively shifted by 250 meV. The interstitial Nb sites will also act as impurity scattering sites, explaining the broader features observed by ARPES here. This is consistent with the high density of atomic-scale imperfections observed in STM measurements on 3R-Nb$_{(1+x)}$S$_2$ \cite{Machida.PRB.2017}, compared with the much cleaner surface of 2H-NbS$_2$ \cite{Guillamon.PRL.2008} and other TMDs. Correspondingly, in transport measurements the residual resistivity ratio for 3R-Nb$_{(1+x)}$S$_2$ samples is vastly inferior to the stoichiometric 2H-NbS$_2$ \cite{Niazi.JPCM.2001}. Thus, taking our results together with the literature, we argue that it is the presence of donor-type interstitials that principally distinguishes the electronic and physical properties of 3R-Nb$_{1+x}$S$_2$ from 2H-Nb$_2$, rather than the difference in stacking arrangement. 

If a stoichiometric 3R-NbS$_2$ existed, our DFT calculations suggest it would have a similar Fermi surface to the 2H phase, and therefore could have similarly interesting properties, potentially including non-centrosymmetric superconductivity. Unfortunately, stoichiometric 3R-NbS$_2$ is entirely hypothetical, and the only thermodynamic bulk phases are 3R-Nb$_{(1+x)}$S$_2$ and 2H-NbS$_2$ \cite{Fisher.Inorg.Chemi.1980}. However, the monolayer limit provides a third structural form  of "1H" NbS$_2$, which has been predicted \cite{Bianco.Nano.Letters.2019} and observed \cite{Lin.Nano.Research.2018} to enter a CDW phase, although this may be substrate dependent \cite{Stan.PRM.2019}, and offers an interesting playground to tune the structural and superconducting instabilities \cite{Devarakonda.Science.2020}. 

\section{Conclusion}

In conclusion, the measured Fermi surfaces of 3R-Nb$_{(1+x)}$S$_2$ are found to be much smaller than in 2H-NbS$_2$, consistent with a large rigid band shift caused by extra charge from interstitial Nb. Thus it is the stoichiometry, rather than the stacking sequence, that principally determines the differences in electronic structure and physical properties between the two polytypes. Our high resolution data on 2H-NbS$_2$ reveals kinks in the spectral function, but the strength of the coupling is found to be much larger for the the sections of bands with Nb 4$d_{x^2-y^2,xy}$ character than for the Nb $4d_{3z^2-r^2}$. Our measurements provide an experimental framework for interpreting the two-gap superconductivity and latent CDW in 2H-NbS$_2$, while also giving insight into the absence of these in the 3R-Nb$_{(1+x)}$S$_2$.    
\\

\begin{acknowledgments}
We thank Niko Tombros and HQ graphene for providing details of sample growth and characterisation. We thank Timur K. Kim for insightful discussions and technical support. We acknowledge Diamond Light Source for time on beamline I05 under proposal CM26443. Z. El Youbi acknowledges Diamond Light Source and CY Cergy-Paris university for PhD studentship support.
\end{acknowledgments}


\begin{thebibliography}{51}
\bibitem [{2H3()}]{2H3Rgeometry}%
	\BibitemOpen
	\href@noop {} {}\bibinfo {note} {{The different stacking between the two
			phases also affects the shape of the Brillouin zone. In the 2H phase it is a
			simple hexagonal prism, however in the 3R phase the geometry of the 3D
			Brillouin zone is more complex, as can be seen in Fig.~2(e). Note also that
			Brillouin zone is twice the height along the $c*$ axis in the 3R phase. For
			convenience of comparison, we continue to use the familar notation of ``K”
			and ``M” also in the 3R phase (e.g. in Fig.~1(f)), even though these are
			not formal high-symmetry points in this case. Similarly, we draw a simple
			hexagonal Brillouin zone in Fig.~2(f,g) that corresponds to a
			surface-projected Brillouin zone.}}\BibitemShut {Stop}%
\end{thebibliography}

%

\end{document}